\documentclass[useAMS]{mn2e_modmargin}
\usepackage{amsmath}
\usepackage{url}
\usepackage{amsfonts}
\usepackage{amsbsy}
\usepackage[dvips]{graphics}
\usepackage{subfigure}
\usepackage{verbatim}
\usepackage{amssymb}
\usepackage{amsbsy}

\renewcommand{\d}{\ensuremath{\partial}}

\newcommand{\ii}{\ensuremath{\text{i}}}

\title[The ballistic transport instability]{The ballistic transport
instability in Saturn's rings II: nonlinear wave dynamics}
\author[Latter, Ogilvie \& Chupeau]{Henrik N. Latter$^{1}$\thanks{E-mail:
    hl278@cam.ac.uk},
   Gordon I. Ogilvie$^{1}$,
   Marie Chupeau$^{1,2}$ \\
$^{1}$ DAMTP, University of Cambridge, CMS, Wilberforce Road,
Cambridge CB3 0WA, UK\\
$^{2}$ LPTMC, Universit\'e Pierre-et-Marie-Curie, Tour 24, 4, Place
Jussieu, 75252, Paris Cedex 05, France}
\date{}

\begin{document}

\maketitle

\begin{abstract}
The ejecta discharged by impacting meteorites can redistribute 
a planetary ring's mass and angular momentum. 
This `ballistic transport' of ring properties instigates a
  linear instability that could generate the 100--1000-km undulations  
  observed in Saturn's inner B-ring and in its C-ring. 
  We present semi-analytic results demonstrating how the instability
  sustains steadily travelling nonlinear wavetrains. 
  At low optical depths, the instability produces
   approximately sinusoidal waves of low amplitude, which we identify with 
   those observed between radii 77,000 and 86,000 km in the
   C-ring. 
  On the other
  hand, optical depths of 1 or more exhibit hysteresis, whereby the ring
  falls into multiple stable states: the homogeneous background
  equilibrium or large-amplitude wave states. 
  Possibly the `flat zones' and `wave zones'
  between radii 93,000 and 98,000 km in the B-ring correspond to the stable
  homogeneous and wave states, respectively. 
  In addition, we test the linear stability of the wavetrains and
  show that only a small subset are stable. In particular,
  stable solutions all possess 
   wavelengths greater than the lengthscale of fastest linear growth. 
  We supplement our calculations with a
  weakly nonlinear analysis that suggests the C-ring 
  reproduces some of the dynamics 
   of the complex Ginzburg--Landau equation. 
  In the third paper in the series, 
 these results will be tested and extended with numerical
  simulations.
\end{abstract}


\section{Introduction}

The component particles of planetary rings suffer a continual
bombardment of hypervelocity meteoroids, the impacts of which
 liberate a
significant amount of material. Typically, impact ejecta
reaccrete on to the ring but at a different radius from where they
originated; they hence redistribute its mass and angular momentum.
This `ballistic transport' of ring properties occurs on
a characteristic lengthscale $l_\text{th}\sim 10-10^3$ km (the `throw
length') and a timescale $t_e \sim 10^5-10^7$ yr (the `erosion time') (Durisen 1984, Ip
1984, Lissauer 1984). 
Other than influencing
the large-scale evolution of Saturn's rings, ballistic transport 
instigates a linear instability that can spontaneously create structure 
on these scales (Durisen 1995). 
It has been argued that the 100-km waves in the inner B-ring and the
1000-km undulations in the C-ring are a result of the instability's
nonlinear saturation (Durisen et al.~1992, hereafter D92, Durisen 1995, Charnoz
et al.~2009, Colwell et al.~2009). 

This is the second paper in a series devoted to the dynamics of the
ballistic transport instability (BTI) and its generation of
axisymmetric structure. The first paper, 
Latter et al.~(2012) (hereafter Paper 1), outlined a convenient
theoretical framework within which to attack the problem and
 rederived the BTI's linear theory.
Here we aim to go further by tracking
the BTI's nonlinear saturation. Ultimately, one is
obliged to numerically simulate its evolution, and we
present such calculations in the third paper of the series (Latter et
al.~2013, submitted, hereafter Paper 3). 
In this work, however, we take a dynamical systems approach and establish
a set of `a priori' results that can both guide and explain
the simulations. 

First we demonstrate that ballistic transport
supports families of
steadily travelling nonlinear wavetrains. 
These solutions may be computed directly
from the system's governing evolution equation. 
At low optical depths $\tau$,
 the wavetrains
assume small amplitudes and possess 
approximately sinusoidal profiles.We identify them
with the long 1000-km undulations in the C-ring between radii 
77,000 and 86,000 km (see Fig.~13.17 in Colwell et al.~2009), 
but conclude that the
100-km plateaus at slightly larger radii are not generated by the BTI, at least 
not working in isolation.
Meanwhile, when $\tau\gtrsim 1$ the system exhibits
hysteresis: the homogeneous state is linearly
stable, but there exist additional wave solutions of large
amplitude. This raises the possibility that stable homogeneous
states spatially adjoin stable wave states, with the interfaces
possibly undergoing their
own dynamics.This theoretical scenario compares well with
observations of `flat' and `wave' zones in the inner B-ring between
radii 93,000 and 98,000 km (see Fig.~13.13 in Colwell
et al.~2009). For sufficiently small viscosities, hysteresis
 extends to very large optical depths. In fact, one can find 
nonlinear BTI-supported waves for $\tau > 2.5$, though it is unlikely 
such structures are relevant to ring observations.

Subsequently, we determine the linear stability of 
these solutions and find that 
  only a
  small subset are stable. As stable solutions possess
 wavelengths longer
  than that of the fastest growing linear mode, it is 
  likely
  that the system
  undergoes a wavelength selection process, whereby power initially
  localised to the most unstable lengthscale 
  seeks out the longer stable wavetrain solutions.
Finally, we conduct a weakly nonlinear analysis of the long and slow
dynamics of wavetrain modulations. It turns out that the wave amplitudes
obey the complex Ginzburg--Landau equation, which suggests that the C-ring
undulations share some of its non-trivial dynamics.

The structure of the paper is as follows. In the following section, we
 summarise the relevant contents of Paper 1, such as the governing
mathematical formalism, main parameters, and the BTI's linear
stability analysis. In Section 3 we calculate the nonlinear
wavetrain solutions, focussing on the two parameter regimes associated
with the C-ring and inner B-ring. Section 4 outlines the linear
stability of these structures, while Section 5 and the Appendix
present a weakly nonlinear analysis of their long and slow
modulations. We bring together these various results in the final
Discussion section.

\section{Preliminaries}

In this section we present relevant background material:
the evolution equation for the dynamical optical depth under
the influence of ballistic transport and viscous diffusion, its key
functions and parameters, and the linear theory of the BTI. 
The presentation is brief and includes no
derivations; more details can be found in Paper 1. 

\subsection{Physical and mathematical formalism}

We employ a local model, the shearing sheet, which describes 
the dynamics of a small patch of ring. 
Doing so means we omit large-scale features such
as edges, and gradients in ring properties, but the model
does isolate cleanly the intrinsic behaviour of the BTI.
The time evolution of the optical depth
is determined by the mass conservation equation, which 
can be cast in the following
dimensionless form:
\begin{align} \label{GovEq}
\d_t \tau = \mathcal{I}-\mathcal{J} +
\tfrac{1}{2}\d_x\left(\mathcal{K}+\mathcal{L}\right) + \mu \d_x^2 \tau.
\end{align} 
Here, $\tau$ denotes dynamical optical depth (dimensionless surface
density) 
defined through $\tau=\sigma/\sigma_1$,
where $\sigma$ is surface density and $\sigma_1$ is the reference
surface density
associated with $\tau= 1$ (see Section 2.4 in Paper 1), 
 $x$ is the radial coordinate in the shearing sheet,
 and $\mu$ is a (constant) measure of the relative strength of viscous
transport over ballistic transport. In Paper 1, we argued that $\mu$ takes
values $\sim 0.01$ in both the inner B- and C-ring. We usually set it
to 0.025. The nonlinear integral operators
$\mathcal{I}$ and $\mathcal{J}$ account for the direct transfer of
mass by ballistic processes, while $\mathcal{K}$ and $\mathcal{L}$
account for the transfer of angular momentum. The sum of the latter
two is, in fact, proportional to the radial mass flux induced by
the ballistic transport of angular momentum. Finally, 
the units of time and space have been chosen so that the
characteristic ballistic throw length $l_\text{th}$ and the
characteristic ballistic erosion time $t_e$ have been set to 1. From
 Durisen (1995) estimates of these scales are
\begin{align}
l_\text{th} &= 2\times 10^2
\left(\frac{v_e}{10\,\text{m}\,\text{s}^{-1}} \right)
\left(\frac{r_0}{10^5\,\text{km}}\right)^{3/2}\,\text{km}, \\
t_e &= 10^6\left( \frac{10^4}{Y}\right)
\left(\frac{\dot{\sigma}_\text{ref}}{\dot{\sigma}_m}\right)
\left(\frac{\sigma}{100\,\text{g}\,\text{cm}^{-2}}\right)\,\text{yr},
\end{align}
where $v_e$ is the mean ejection speed of ejecta (a sensitive function
of the ring-particle surface), $r_0$ is the radius, $Y$ is the yield (the ratio of ejecta mass to
the mass of the impacting meteoroid), $\dot{\sigma}_m$ is the
one-sided meteoroid flux at Saturn, and the reference flux is
$\dot{\sigma}_\text{ref}=4.5\times 10^{-17}\,\text{g}\,\text{cm}^{-2}\,\text{s}^{-1}$.

The integral operators involve three important functions: the rate of
ejecta emission per unit time and area $R(\tau)$, the probability of
mass absorption from incoming ejecta $P(\tau)$, and the ejecta distribution
function $f(x)$, defined so that $f(x)dx$ is the proportion of
material thrown distances between $x$ and $x+dx$. More generally, $P$
is a function of both the optical depth at the absorbing radius and at
the emitting radius. However, in this paper we deal with the
simpler case when it is only a function of the absorbing radius; this
approximation works best for larger optical depths.
The integral operators in \eqref{GovEq} can be expressed through
\begin{align}
&\mathcal{I} = P[\tau(x)]\int R[\tau(x-\xi)]\,f(\xi)\,d\xi, \\
&\mathcal{J} = R[\tau(x)]\int P[\tau(x+\xi)]\,f(\xi)\,d\xi, \\
&\mathcal{K} = P[\tau(x)]\int \xi\,R[\tau(x-\xi)]\,f(\xi)\,d\xi, \\
&\mathcal{L} = R[\tau(x)]\int \xi\, P[\tau(x+\xi)]\,f(\xi)\,d\xi,
\end{align}
where the integration limits extend from $-\infty$ to $\infty$. 
Note that the integrals may be written as convolutions, a convenience that
facilitates both our analytic and numerical calculations.

The functional
forms for ejecta emission $R$ and absorption $P$ are taken from Paper 1:
\begin{align}
&P(\tau)= 1-\text{exp}(-\tau/\tau_p), \label{P}\\
&R(\tau) = 0.933\left[ 1 + \left(\frac{\tau}{\tau_\mathrm{s}}-1\right)
\exp(-\tau/\tau_\mathrm{s})\right], \label{R}
\end{align}
Following Durisen (1995), we fix the parameters so that the reference optical
depths are $\tau_p=0.5$ and
$\tau_s=0.28$. The throw distribution is
approximated by an off-centred Gaussian profile: 
\begin{align}
f(\xi) = \frac{1}{\sqrt{2\pi d^2}}\,\text{exp}\left[-(\xi-\xi_0)^2/(2d^2)
\right]. \label{f}
\end{align}
To best match with Cuzzi \& Durisen (1990) we set the off-set to $\xi_0=0.5$ and the standard
deviation to $d=0.6$. There is a case for varying the parameters
$\tau_p$ and $\tau_s$ in different regions of the ring, and indeed
$f$ may differ significantly in the peaks of C-ring plateaus where ring
particles are smaller and easier to destroy (Estrada and Durisen
2010). However, these complexities 
obscure the most important dynamics, and are not pursued in this
paper. 
Indeed, we anticipate they bring only minor
qualitative changes to the BTI's evolution, and perhaps
only minor quantitative changes as well -- at least within the
many uncertainties. Finally, note that we do not account for the ring
viscosity's dependence on surface density 
(Araki and Tremaine 1985, Wisdom and Tremaine 1988, Daisaka et al.~2001). Again, this
simplifies the analysis, while not changing the qualitative behaviour
of the dynamics.

\subsection{Linear theory of the BTI}

We next outline the main characteristics of the linear BTI.
 Assuming a homogeneous background state
$\tau=\tau_0$, small perturbations of the form $\propto \text{exp}(st + \ii k
x)$ grow according to the dispersion relation
\begin{equation}\label{dispgen}
s = R_0'P_0\,H(k) - R_0P_0'\,\overline{H(k)} - \mu k^2,
\end{equation}
where
\begin{equation}\label{G}
H(k)= F(k)-F(0) - \frac{1}{2}k\left[F'(k)+F'(0)\right],
\end{equation}
and $F(k)$ is the (non-unitary) Fourier transform of $f$. The overbar
denotes the complex conjugate, a prime
indicates differentiation with respect to $\tau$, and a subscript $0$
indicates evaluation at $\tau=\tau_0$. With an off-centred Gaussian
model for the distribution $f$, the expression \eqref{dispgen} may be
evaluated analytically using
\begin{equation}
F(k)= \text{exp}\left(-\tfrac{1}{2}d^2 k^2 - \ii k \xi_0\right).
\end{equation}

\begin{figure}
\begin{center}
\scalebox{0.5}{\includegraphics{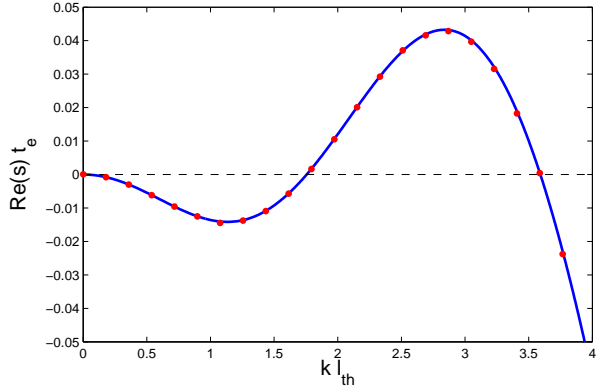}}
 \caption{The solid line indicates the 
 real growth rate $s$ of the ballistic transport
   instability as a function of waveumber $k$,
   in a homogeneous ring of $\tau_0=0.175$ and $\mu=0.025$ (cf. Section
   2.2). The solid points indicate the growth rate of the unstable
   mode that attacks a very low amplitude steady wavetrain with
   $q=3.5874$, travelling over a background of $\tau_0=0.175$ (cf.\
   Section 4).}\label{disp1}
\end{center}
\end{figure}

The main parameters governing the growth of a mode are the
equilibrium optical depth $\tau_0$, the ballistic Prandtl number
$\mu$, and the wavenumber $k$ of the mode. In
Fig.~\ref{disp1} we plot representative growth rates for a low optical
depth ring. See also Fig.~\ref{crit} for a marginal case, in which the
maximum real growth rate is exactly 0 for non-zero $k$. 
The figures show that instability is restricted to an
intermediate range of wavelengths: both very short and very long waves are stable.
Note also that the BTI typically takes the form of a growing travelling
wave, with the direction and speed of propagation controlled by the
asymmetry in the throw distribution $f$.

\begin{figure}
\begin{center}
\scalebox{0.5}{\includegraphics{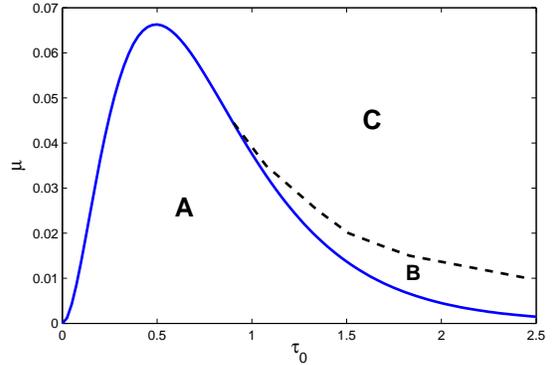}}
 \caption{Curves of marginal stability and hysteresis in the $(\tau_0,\mu)$
   plane. Region `A', enclosed by the solid curve, denotes
   parameter values that are linearly unstable to the BTI. Both regions `B'
   and `C' are linearly stable. In region `B', however,
   large-amplitude steady nonlinear wavetrains exist, even though the linear
   BTI is suppressed.}\label{stab1}
\end{center}
\end{figure}

For the throw
distributions we consider, 
a necessary (but not sufficient) condition for instability  is
\begin{align}
\frac{d \ln P}{d\tau} > \frac{d \ln R}{d \tau},
\end{align}
which states that an overdensity grows if it emits less
material than it can absorb. The functions \eqref{P} and \eqref{R}
ensure 
this condition is satisifed for all
intermediate $\tau_0$. A sufficient condition for instability,
however, must involve viscous damping and hence the parameter
$\mu$. Figure \ref{stab1} presents regions
 of instability and stability 
 in
the parameter space of $\tau_0$ and $\mu$;
Region `A' is unstable, while Regions `B' and `C' are stable.
Given that $\mu$ varies
between 0.01 and 0.05 between the C-ring and inner B-ring (Paper 1), both
ring regions should be near marginal stability. 
In Paper 1, we speculated that this could strongly influence
the BTI's nonlinear development, leading either to low-amplitude
saturation or bistability. These expectations are verified in this
paper. 

\section{Nonlinear wavetrains}

Unstable BTI modes grow exponentially and independently until they
leave the linear regime and begin to interact. At this point we are
normally obliged to pursue
their evolution numerically. Previous
  work (D92, for example) focussed
 on how ballistic
 transport sculpts inner ring edges;
it did not
directly track the BTI, even if it was sometimes present. 
Our numerical simulations in Paper 3, in contrast, mostly dispense with ring edges and concentrate
explicitly on the BTI's nonlinear development in isolation.
In
this paper we explore an
alternative approach to both: 
instead of simulating Eq.~\eqref{GovEq}, we calculate
its exact steady nonlinear solutions. These coherent structures,
understood as fixed points in the system's phase space, 
control the long-time behaviour of the instability, by either repelling
or attracting the system's phase trajectories. As a consequence, they
provide insights into the general dynamics and hence 
the simulation results.

The simplest non-trivial invariant solutions of \eqref{GovEq} take the
form of steadily travelling wavetrains. Indeed, wavetrains appear in the
simulations of D92 for certain parameters.
Moreover, the existence of such travelling 
structures can be anticipated from the
mathematical
structure of the problem, i.e.\ its translational symmetry (within the
local approximation) and the
fact that the onset of instability takes the form of a Hopf
bifurcation. Of course, the strongest indication that the BTI supports waves
are the observations themselves, with wavetrains permeating the inner
B-ring and the C-ring 
(Horn \& Cuzzi 1996, Porco et al.~2005, Colwell et al.~2009).
 
Though the BTI sustains a rich variety of
nonlinear wavetrains, we
do not attempt a comprehensive exploration in this paper.
To bring out the most relevant results, we
restrict the analysis to parameter regimes corresponding to the C-ring
and the
inner B-ring. Each is treated separately in
subsections 3.2 and 3.3, before a more general discussion.
First, however, we present our mathematical and numerical approach.

\subsection{Nonlinear eigenvalue problem}

To calculate travelling solutions to \eqref{GovEq}
 a co-moving spatial co-ordinate $\zeta$ is introduced, defined so that
\begin{align} 
\zeta= x - c_p t, \label{comovy}
\end{align}
where $c_p$ is the phase speed of the wave.
To capture wavetrain solutions, 
 we assume $\tau$ is periodic in
$\zeta$ with wavelength $\lambda=2\pi/q$, where $q$ is a specified
constant wavenumber. As a consequence, 
Eq.~\eqref{GovEq} transforms into an
ordinary integro-differential equation for $\tau$ in terms of $\zeta$, which we write as
\begin{align} \label{wavetrains}
c_p \d_\zeta \tau +
\mathcal{I}-\mathcal{J}+\tfrac{1}{2}\d_\zeta(\mathcal{K}+\mathcal{L})
+ \mu\,\d_\zeta^2 \tau=0.
\end{align}
The integral operators in the above can be recast in a
straightforward way, keeping the integration limits $\pm \infty$. 

Equation \eqref{wavetrains} describes a nonlinear
eigenvalue problem for $\tau$ with eigenvalue $c_p$. We must apply
periodic boundary conditions, so that
$\tau(\zeta+\lambda)=\tau(\zeta)$.
An additional constraint
 is the preservation of a specified mean optical
depth $\tau_0$,
i.e.
\begin{equation} \label{constraint}
\tau_0 = \frac{1}{\lambda}\int_0^\lambda \tau\, d\zeta.
\end{equation}
The three parameters governing the problem are hence $\tau_0$, $\mu$,
and the wavenumber $q$. Recall that we keep the parameters appearing
in our definitions of $R$, $P$, and $f$ constant throughout this paper.

The problem is solved using a Fourier pseudo-spectral method. The
domain $0 \le \zeta<\lambda$ is split into $N$ equal components. Typically we
set $N=512$, which supplies excellent convergence. 
The $\zeta$ derivatives in \eqref{wavetrains} are computed by an
excursion into spectral space with a FFT. The $\zeta$ integrals are also
evaluated in spectral space, using the convolution theorem.  
Equations \eqref{wavetrains} and \eqref{constraint} then become $N+1$ nonlinear
equations for $N+1$ unknowns: $c_p$ and the values of $\tau$ at the
$N$ grid points. These are solved by a multidimensional 
Newton-Raphson algorithm.

\subsection{The C-ring: low amplitude wavetrains}

This subsection examines the low-$\tau$ regime relevant to Saturn's
C-ring. We adopt fiducial parameters of $\mu=0.025$ and
 $\tau_0= 0.175$. For this choice
our numerical method uncovers a family of nonlinear wavetrains of
small but non-negligible amplitudes in an interval of $q$ between
roughly 1.75 and 3.59. These limiting values coincide with the
wavenumbers of the two marginal linear modes for
which Re($s)=0$, as shown in Fig.~\ref{disp1}. Thus the marginal
 linear modes
bracket the wavelengths of allowed nonlinear waves.

\begin{figure}
\begin{center}
\scalebox{0.45}{\includegraphics{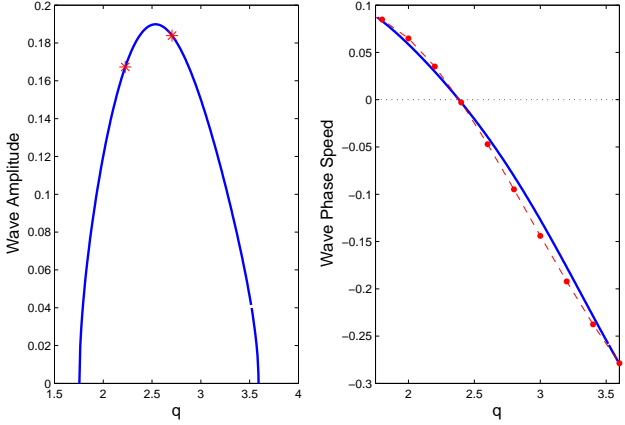}}
 \caption{Solid curves correspond to 
 the amplitudes and phase speeds $c_p$ of nonlinear
 wavetrains as functions of wavenumber $q$ for C-ring parameters:
$\tau_0=0.175$ and $\mu=0.025$. In the second plot the wavespeeds of 
the linear BT modes are also plotted as data points joined by a dashed
curve. Wavetrains between the two asterisks in the first panel are
linearly stable (cf.\ Section 4).}
\label{Cringvsq}
\end{center}
\end{figure}

We measure the amplitudes of the wavetrains with 
$\text{max}(\tau)-\tau_0$. In the left panel of 
Fig.~\ref{Cringvsq}
we plot the amplitude as a function of $q$. 
Its maximum value is a little less than $0.2$ and occurs for $q\approx 2.54$,
less than the wavenumber of fastest linear growth, which is
closer to 2.84.
In the second panel the solid curve represents the corresponding
phase speeds $c_p$. 
Interestingly, the $q$ at which $c_p=0$ is approximately that
which yields the largest wave amplitude. 
Superimposed, as data points, are the
wavespeeds of the \emph{linear} BTI modes. The two curves are
very similar, sharing the same values at the endpoints of the $q$
interval. Finally, the group velocity, $c_g= d(qc_p)/dq$, is
negative for all permitted $q$: information always 
travels inwards.
This is true even for longer waves for 
which the wave crests, in contrast, 
travel outwards ($c_p>0$). Generally $|c_g|>|c_p|$ and varies between
-0.192 and -1.13,
with the longest waves possessing the slowest group velocities.

\begin{figure}
\begin{center}
\scalebox{0.5}{\includegraphics{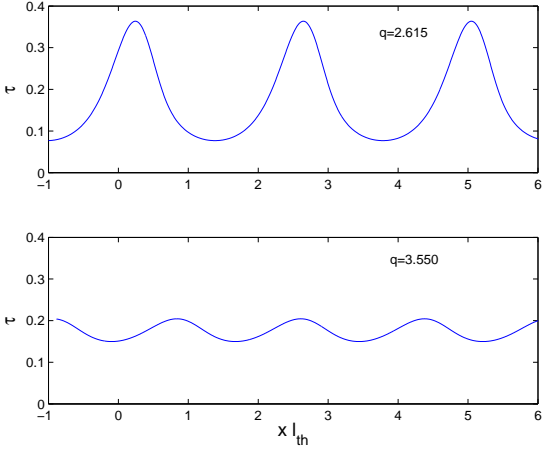}}
 \caption{Profiles of two wavetrains corresponding to $q=2.615$ (top
   panel) and $q=3.550$ (bottom panel). The mean optical depth
   $\tau_0$ is
   $0.175$ and $\mu=0.025$. }\label{Cringprofs}
\end{center}
\end{figure}

In Fig.~\ref{Cringprofs} we present two representative
wavetrain profiles. The top panel corresponds to a comparatively 
large-amplitude wave with $q= 2.615$. The variation in $\tau$ between peak
and trough is roughly 0.2, similar to what is observed in the C-ring. 
We plot a comparatively low-amplitude wave in the lower panel with a
shorter wavelength. Here $q=3.550$, meaning the wavetrain is very near the
limiting $q$ beyond which no solutions exist. As a consequence,
it is essentially the same as the marginal (and steady) linear BTI
mode, and thus exhibits a sinusoidal profile.

These results suggest that the BTI when near marginality, as
it is in the C-ring, tends to saturate in low-amplitude
travelling waves. These wavetrains possess a wavelength $\lambda$ between
approximately 1.5 and 3 throw lengths $l_\text{th}$, values which also
bracket the set of unstable linear modes. Given their general
sinusoidal appearance and the fair correspondence between the linear
and nonlinear wave speeds (Fig.~\ref{Cringvsq}b), they
invite a weakly nonlinear analysis, which yields
their saturation
amplitudes analytically. The calculation is outlined
in Section 5 and the Appendix. There it is
also shown that, in addition to wavetrain solutions, there also exist
solutions that consist of long travelling modulations of these same
wavetrains.

Our solutions probably correspond to the `ripples' that
appear in low-$\tau$ regions in some D92 simulations. Though those
authors conjecture that the waves are driven by the B-ring edge, they
also leave open the idea that they could be generated by an instability
working in isolation, which
is what we show here.
 
Because the characteristic throw length $l_\text{th}$ is poorly
constrained, it is difficult to unambiguously compare ballistic transport
results with the observed features in the C-ring.
 Do our solutions correspond to the 100-km plateaus or
the 1000-km undulations?
 Given the two markedly different lengthscales of these features, and
the fact that they can occur at the same radii, it is
unlikely that the BTI generates both concurrently.
We associate the BTI with the 1000-km
 undulations. The general morphology of the theoretical profiles 
(low amplitude and generally sinusoidal) bears a closer
resemblance to the long undulations than to the
 plateaus and their characteristic `flat-top' profile. 
A
consequence of this identification is that $l_\text{th}\sim 500$ km,
at least in the C-ring.

\subsection{The B-ring: hysteresis}

In this subsection we adopt a parameter regime corresponding
to the inner B-ring. We take $\mu=0.025$ again and set $\tau_0=1.3$.
 This choice is suitable for a situation amidst a
`wavy' zone, rather than a `flat' zone, in which the mean optical depth
is slightly less (Colwell et al.~2009).

For these parameters the linear theory states there are no growing
modes. The BTI is extinguished because the homogeneous state is
too optically thick. When $\mu=0.025$,
the largest $\tau_0$ that supports instability is $\tau_0\approx
1.2$. Nevertheless, we are able to compute
 wavetrain solutions of
non-trivial amplitudes. Similarly to lower $\tau_0$, they occur in a finite
interval of wavenumber: $q$ must lie between approximately 2 and 3. 
Moreover, we find two distinct families of solutions.
For fixed $q$ there exist two wavetrains of differing amplitude and
morphology. 
Note that as the stable homogeneous state $\tau=1.3$ is also a
solution for these parameters, the system is potentially
 bistable. 

\begin{figure}
\begin{center}
\scalebox{0.45}{\includegraphics{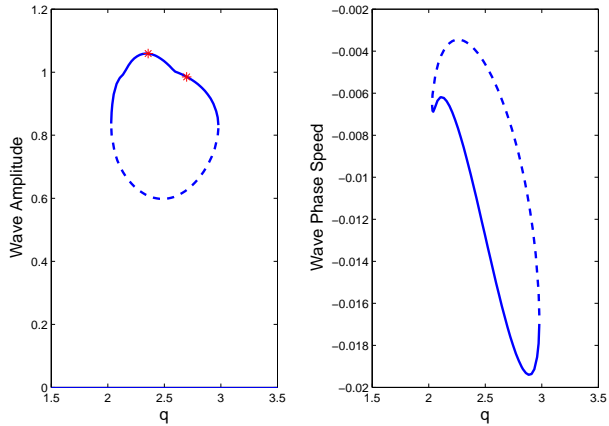}}
 \caption{The amplitudes and phase speeds $c_p$ of wavetrain solutions as functions of
   wavenumber $q$ for parameters corresponding to the inner B-ring:
   $\tau_0=1.3$ and $\mu=0.025$. There are two branches of wave
   solutions which we
   distinguish by solid and dashed lines. In addition, the
   homogeneous state $\tau=1.3$ is also a solution. 
  Upper branch solutions lying between the two asterisks are linearly
  stable. All the members of the lower branch are unstable (cf.\
  Section 4). There also exists a very narrow band of stable upper
  branch solutions near $q=2.04$ which is not indicated by asterisks.}\label{Bringvsq}
\end{center}
\end{figure}

The solution branches are plotted in the left panel of
Fig.~\ref{Bringvsq} as functions of wavenumber $q$. The
right panel shows the associated phase speeds $c_p$. The solid line
indicates the upper branch of solutions, and the dashed line describes
the lower branch. Taken together the two families 
form an `isola' in the solution
space. Unlike the low
$\tau_0$ case explored earlier, the
amplitudes of the wavetrains are large, and the
wavecrests propagate inwards, albeit extremely slowly. For fiducial values
of $l_\text{th}$ and $t_e$, the phase speed lies between 1 and
100 mm/yr. We certainly expect no measurable difference in
the wave
positions since the first Voyager images of the B-ring.  
The group velocity $c_g$, though typically negative, can
take positive values near the upper and lower limits of permitted $q$.

\begin{figure}
\begin{center}
\scalebox{0.4}{\includegraphics{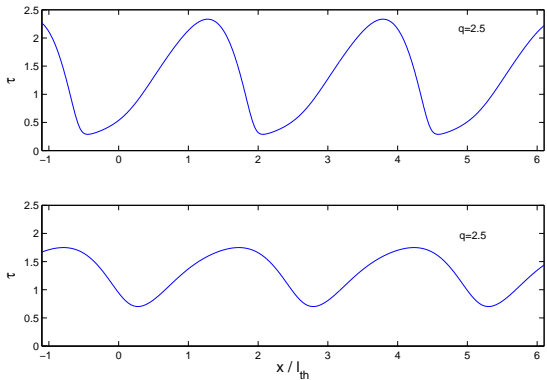}}
 \caption{Two representative examples of wavetrain solutions for
   $\tau_0=1.3$ and the same wavenumber $q=2.5$. 
   The top panel shows the stable wavetrain from the upper branch of
   solutions, while the bottom panel shows the unstable 
   one from the lower branch.}
  \label{BringProf}
\end{center}
\end{figure}

 In Fig.~\ref{BringProf} we present two
examples of the solutions' profiles. The top panel
corresponds to a wavetrain from the upper wave branch, and the bottom
panel from the lower branch. Note that $\tau$ in the
upper branch waves varies between roughly
0.4 and 2.3, from trough to peak. 
In contrast, the lower branch exhibits a more narrow
variation. Both profiles deviate appreciably from the
sinusoidal shapes of the low-$\tau_0$ solutions of Section 3.2. 
The upper branch solution exhibits slight `ramp-like'
features to the right of its minima,
 a feature that
 becomes
 more evident at larger wave amplitude (see next subsection).      
The morphology of the upper branch matches fairly well with
those produced by Durisen et al.'s simulations of the inner B-ring
edge (D92); see for example their
Fig.~6. Consequently, we regard the D92 waves as direct analogues of
our solutions.

From the structure of the solution space, we anticipate that the lower
wave branch is linearly unstable. The system will prefer to migrate to
either the `flat' homogeneous state or one of the upper `wave'
states. We show this in some detail in Section 4. 
This, however, causes trouble
when we compare the morphologies of the stable
upper-branch waves
with the observed waves in the inner B-ring. 
As mentioned, the former exhibit very
deep troughs ($\tau \sim 0.4$) and large peaks ($\tau \sim 2.3$), 
while the latter's optical depth variation is less marked, with $\tau$
ranging between roughly 0.8 and 1.9 (Colwell et al.~2009). 
Inconveniently, the unstable lower
branch waves offer a much better fit! Of course, our theoretical
profiles are framed in terms of \emph{dynamical} optical depth, 
while the Cassini cameras
measure \emph{photometric} optical depth, and the likely presence of self-gravity
wakes ensures the two quantities differ.
It is unclear, though, if this can fully account for the
discrepancy: obviously, further work is needed. 
Finally, it is worth mentioning that
the more detailed Durisen et
al.\ calculations also share the same deep troughs (D92), and so it is
unlikely that the disagreement arises because of
idealisations in our model.

Because two stable states are possible at any given location
 the spatial domain may split up into `flat' and `wave' zones,
 each separated by a `front' that may itself move, but at a speed different
 from $c_p$. Such a partitioning is indeed what the observations show.
In principle, it is possible to analytically explore families of such
`homoclinic' structures (Burke and Knobloch 2007). Unfortunately
 our system exhibits wave zones that spread as well as travel, and as
 a consequence are difficult to work with. They are more easily
 treated via numerical simulations, and we show detailed examples
 in Paper 3.

\subsection{General structure of solution space}

Having examined two representative examples in detail, 
we now
summarise the general solution structure as
$\tau_0$ and $\mu$ vary, in addition to the wavenumber $q$. We, however,
give slightly more emphasis to the $\tau_0$ dependence
 because the solutions' $\mu$
dependence is less interesting. 

\begin{figure}
\begin{center}
\scalebox{0.5}{\includegraphics{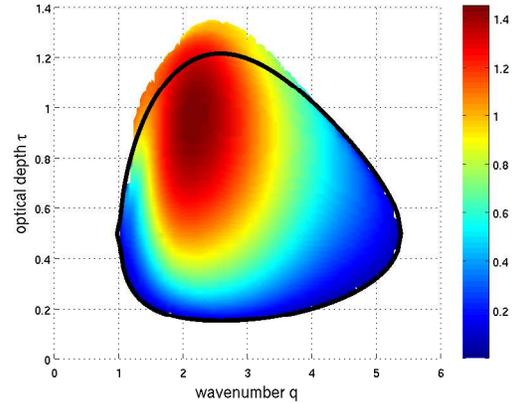}}
 \caption{The amplitude of wavetrain solutions as a function of
   $\tau_0$ and $q$ for fixed $\mu=0.025$. White regions indicate that
 no wave solutions exist. The thick black line encompasses the region
 in parameter space that is linearly unstable. As wave solutions exist
 in regions that are linearly stable the system exhibits hysteresis. In
 bistable regions only the amplitude of the upper wave branch is
 plotted.}
\label{SolnSurf}
\end{center}
\end{figure}

\begin{figure}
\begin{center}
\scalebox{0.5}{\includegraphics{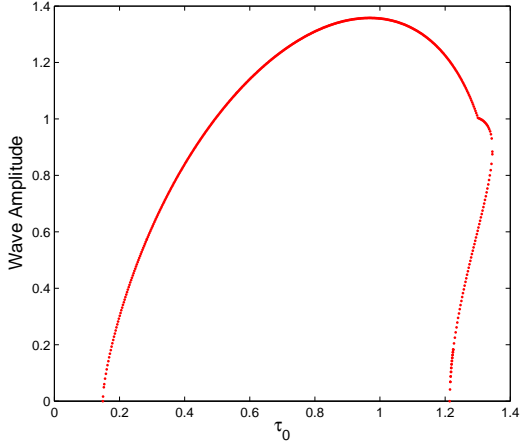}}
 \caption{The wave amplitude as a function of $\tau_0$ for
   fixed $q=2.5$ and $\mu=0.025$. Both upper and lower wave branches
   are included. Between roughly $\tau_0=1.2$ and $1.35$ the system
   exhibits hysteresis.}
\label{Hyst-tau}
\end{center}
\end{figure}

Figure \ref{SolnSurf} presents coloured contours of the wave amplitude
as a function of $q$ and $\tau_0$, with $\mu=0.025$. 
The white area indicates where no
wave solutions exist. The thick black curve
encompasses the region within which linear modes grow. For most values
of $\tau_0$ and $q$, wave solutions are confined within the linear
curve. But when $\tau_0$ is larger there exist solutions for parameters
where the homogeneous state is stable. These regions 
exhibit bistability and admit two wave solutions with
different amplitudes (as in Section 3.3), though we only plot the upper branch
amplitudes
 in
Fig.~\ref{SolnSurf}.

The maximum amplitude for a given $\tau_0$ always occurs
at a $q$ smaller than that of the fastest growing linear
mode. Furthermore, the $q$ that yields the
maximum amplitude and the $q$ that gives $c_p=0$ are relatively close to
each other. 
For example, at $\tau_0=1$, the former $q$ is $2.18$ and the latter
is $2.32$, while the fastest growing linear mode possesses
$q=3.02$. When $\tau=0.5$, the largest amplitude occurs at $q=2.12$, while
$c_p=0$ occurs at $q=2.42$, and the fastest growing mode has
$q=3.45$. 

Near $\tau_0 \approx 0.9$
and for $q\approx 1.3$ the solution surface is complicated and appears
to `tear', with nearby regions twisting upwards to either larger amplitudes
or downwards to zero amplitude. This suggests there may be additional solution
branches. We have not attempted to compute these
additional 
(hypothetical) longer wavelength structures, and have not
observed them in the simulations of Paper 3.   

In Fig.~\ref{Hyst-tau} we plot the wave amplitude as a function of
$\tau_0$, keeping both $q$ and $\mu$ constant. Both upper and lower solution branches
are included and thus the figure clearly represents the
hysteresis at larger $\tau_0$.  When $\mu=0.025$, 
hysteresis occurs in a relatively small region
in parameter space, between roughly $\tau_0=1.2$ and 1.35. Notable is
the large amplitude of the upper wave state; this could
mean that large disturbances are needed to transfer a portion of ring
from the flat state to the wave state and vice versa. 
Perhaps the Janus/Epimetheus 2:1 inner
Lindblad resonance, which falls within a wave region in the inner
B-ring,
 could provide such a strong disturbance (Colwell et al.~2009).

\begin{figure}
\begin{center}
\scalebox{0.5}{\includegraphics{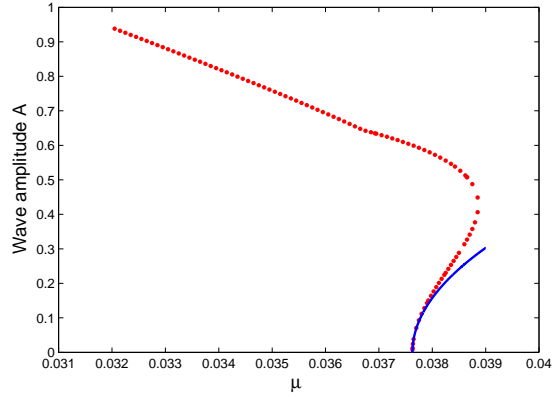}}
 \caption{The red points represent the numerically computed amplitude as a function of $\mu$ for
   fixed $q=2.57453$ and $\tau_0=1$. Both upper and lower wave branches
   are included. The solid blue line represents amplitudes calculated from
   the weakly nonlinear analysis of Section 5 and the Appendix.}\label{Hyst-mu}
\end{center}
\end{figure}

We explore the dependence of the solutions on $\mu$ in
Fig.~\ref{Hyst-mu}.
There we plot amplitude versus $\mu$ while keeping $\tau_0$ and $q$
constant. Hysteresis is also observed near the critical $\mu$, but
only at
larger $\tau_0$, not at lower $\tau_0$. 
We also plot the amplitude computed from the
weakly nonlinear analysis of the Appendix, namely Equation
\eqref{amp}. This solution is discussed in Section 5.

Lastly, we examine hysteresis in the
 $\tau_0$---$\mu$ plane.
 In Figure \ref{stab1}, it is localised to Region `B'. 
 The dashed curve is achieved by optimising the
critical $\tau$ upon which the solution branches terminate, as $q$ varies but
$\mu$ remains fixed. As is clear from the plot, hysteresis only occurs for higher optical
depths, approximately $\tau > 1$. But what is striking is how far the nonlinear
solutions survive into the linearly stable high-$\tau$ regime. For
$\mu=0.01$, the linear stability shuts down at $\tau_0=1.64$, but nonlinear
waves persist up to $\tau_0=2.45$.
For smaller $\mu$, BTI wavetrains occur at extremely large optical
depths indeed. 
It is improbable, however, that the central and
outer B-ring ---
the only venues exhibiting such high $\tau$ ---
possess $\mu<0.01$ (see discussion in Paper 1). Consequently,
the linear or nonlinear BTI should not play a role there. 

\begin{figure}
\begin{center}
\scalebox{0.4}{\includegraphics{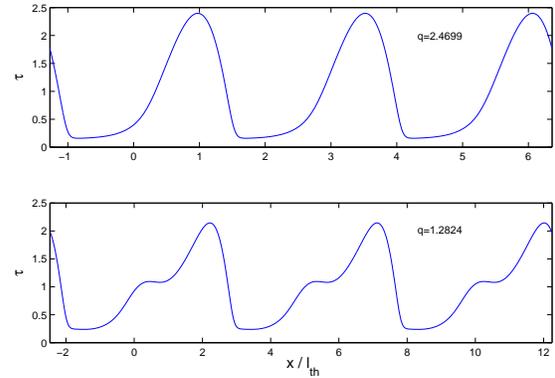}}
 \caption{Two waveforms for $\tau_0=1$ and $\mu=0.025$. The top panel
   shows a wave possessing $q=2.4699$ and 
the lower panel a longer wave with $q=1.2824$.}\label{waveprofs3}
\end{center}
\end{figure}

Before moving on, we show two additional waveforms from different
locations in Fig.~\ref{SolnSurf}. The upper panel of
Fig.~\ref{waveprofs3} shows the largest amplitude wavetrain possible, occurring
for $\tau_0=1$ and $q\approx 2.47$. Its general morphology is similar to
Fig.~\ref{BringProf}a, but it exhibits longer
and deeper troughs, with $\tau$ dipping below $0.2$, as well as more
conspicuous ramp features. 
In the lower panel
we plot a long wavelength wavetrain from the region to the left of
the dark line in Fig.~\ref{SolnSurf} at $\tau_0=1$. This wave
possesses $q\approx 1.28 $. Its morphology is striking, consisting of
a trough, plateau, and peak. Though interesting, it is difficult to
connect this waveform with observations. Moreover,
as we find in Section 4, such long wavelength waves are unstable and unlikely
to play a role in the main dynamics.

\section{Linear stability}

Of all the previously computed wavetrains we expect the linearly stable ones
to dominate the
BTI's nonlinear evolution. Linearly stable solutions
serve as attractors in the system's phase space: the ring is likely to settle
on or around them, and hence exhibit their chief characteristics. 
In this section we determine the stability of the
solutions computed in Section 3.
 Our main result
is that, of
all the various wavetrain solutions available, only a
small subset are actually stable. For given $\tau_0$ and
$\mu$, stable wavetrains usually occur on a narrow band of $q$
encompassing the values that yield $c_p=0$ and the maximum wave amplitude.

\subsection{Modal analysis}

First we set up the mathematical framework with which to determine
stability. The underlying wavetrain is denoted, as before, by $\tau$
and a small perturbation on top of this solution by $\tau'$. 
The system is transferred to a comoving coordinate system
with spatial variable $\zeta$, as defined in Eq.~\eqref{comovy}. 
Once we approximate \eqref{GovEq} for small
$\tau'$, we obtain a linear equation for $\tau'$ in $t$ and $\zeta$ that is
$\lambda$-periodic in $\zeta$. As a consequence, we make the Floquet ansatz
and let $\tau'$ take the following form:
\begin{align}
\tau' = \text{e}^{st + \ii k\zeta} \,\hat{\tau}(\zeta),
\end{align} 
where $s$ is a (complex) growth rate, $k$ is the (real) wavenumber of
the disturbance envelope, the
Floquet exponent, and $\hat{\tau}$ is a $\lambda$-periodic function in
$\zeta$. Recall that $\lambda=2\pi/q$. 
Generally, $s$ and $k$ differ from the linear growth rates that
appear in Section 2.2, though in the limit in which the wavetrain
amplitude $\to 0$ they
do coincide. We need only examine values of $k$ between 0 and
$q/2$; outside this range the solutions repeat.

The governing linearised equation for $\hat{\tau}$ is
\begin{align} 
s\,\hat{\tau} & = \mathcal{I}' -\mathcal{J}' + \tfrac{1}{2}\left(
  \d_\zeta+\ii k \right)\left[\mathcal{K}'+\mathcal{L}'\right] \notag \\
& \hskip3cm+ \mu\left( \d_\zeta^2 +2\ii k\d_\zeta -k^2  \right)\hat{\tau}. \label{nlwlin}
\end{align}
Using periodic summation, the four primed integral operators can be manipulated into the
following forms, 
which are better suited to our numerical method:
\begin{align}
\mathcal{I}' &= P'(\zeta)\,\hat{\tau}(\zeta)\,\int_0^{\lambda} R(\zeta-\xi)
f_\Sigma(0,\xi)\,d\xi \notag \\ 
& \hskip1cm+ P(\zeta)\int_0^\lambda
R'(\zeta-\xi)\,\hat{\tau}(\zeta-\xi)\,f_\Sigma(k,\xi)\,d\xi, \\
\mathcal{J}'&= R'(\zeta)\,\hat{\tau}(\zeta)\,\int_0^{\lambda} P(\zeta+\xi)
f_\Sigma(0,\xi)\,d\xi \notag \\ 
& \hskip1cm+ R(\zeta)\int_0^\lambda
P'(\zeta+\xi)\,\hat{\tau}(\zeta+\xi)\,f_\Sigma(-k,\xi)\,d\xi, \\
\mathcal{K}' &= - P'(\zeta)\,\hat{\tau}(\zeta)\,\int_0^{\lambda} R(\zeta-\xi)
\d_kf_\Sigma(0,\xi)\,d\xi \notag \\ 
& \hskip0.7cm - P(\zeta)\int_0^\lambda
R'(\zeta-\xi)\,\hat{\tau}(\zeta-\xi)\,\d_kf_\Sigma(k,\xi)\,d\xi, \\
\mathcal{L}' &= R'(\zeta)\,\hat{\tau}(\zeta)\,\int_0^{\lambda} P(\zeta+\xi)
\d_kf_\Sigma(0,\xi)\,d\xi \notag \\ 
& \hskip0.7cm+ R(\zeta)\int_0^\lambda
P'(\zeta+\xi)\,\hat{\tau}(\zeta+\xi)\,\d_kf_\Sigma(-k,\xi)\,d\xi.
\end{align}
To ease the notation in the above, we have set $P(\zeta)=P[\tau(\zeta)]$,
$P(\zeta-\xi)=P[\tau(\zeta-\xi)]$, $P'(\zeta)=P'[\tau(\zeta)]$, etc. 
We have also introduced the $\lambda$-periodic
distribution function $f_\Sigma$, defined via
\begin{align} \label{fsum}
f_\Sigma(k,\xi) = \sum_{n=-\infty}^\infty \text{e}^{-\ii k(\xi+ \lambda n)}
f\left(\xi + \lambda n\right).
\end{align}
For the off-centred Gaussian profile of Eq.~\eqref{f}, the new
distribution function can be re-expressed as
\begin{align*}
f_\Sigma = \frac{1}{2\pi} \text{e}^{-\ii\xi_0 k - \frac{1}{2}d^2k^2} 
\Theta_3\left(\tfrac{1}{2}\ii k/q+\tfrac{1}{2}q(\xi-\xi_0),\,\text{e}^{-\frac{1}{2}d^2q^2}\right),
\end{align*}
where $\Theta_n$ is the Jacobi theta function (Whitaker and Watson
1990). However, given the rapid convergence of the series in
\eqref{fsum}, it is more convenient in practice
 to use a truncated series expression
for $f_\Sigma$. 

Equation \eqref{nlwlin} is a linear eigenvalue problem for
$\hat{\tau}$ with eigenvalue $s$. The main parameters comprise
$\tau_0$, $\mu$, and $q$, which specify the nonlinear wavetrain whose
stability we test, and $k$ the wavenumber of the linear mode attacking
the wavetrain.

Because there are multiple modes that are potentially unstable we seek
a numerical method that can retrieve more than one eigensolution at a
time. We hence transform \eqref{nlwlin} into an algebraic eigenvalue
problem by approximating the operator on the right side of the
equation as a matrix. The variable $\zeta$ is discretised on the domain
$[0,\,\lambda]$ into $N$ equally spaced points, and the spatial derivatives are represented by
pseudo-spectral matrices (see Boyd 2002). We approximate by quadrature formulae
the integrals with $\hat{\tau}$ in the integrand. Because the
integrands are periodic, the trapezoidal rule offers spectral
accuracy. We may write such integrals as finite sums, and
hence as matrices
operating on the discretised $\hat{\tau}$. Once the operator on the right side of
\eqref{nlwlin} is reduced to an $N$-by-$N$ matrix, we extract the
eigenvalues and eigenvectors using either the QR algorithm or an
Arnoldi method (Golub and van Loan 1996).

\subsection{Numerical results}

We do not give an exhaustive stability analysis of all wave
solutions; instead we focus on the wavetrains associated with
the C-ring and B-ring, as explored in Sections 3.2 and
3.3.

We first check the stability of extremely low amplitude
wavetrains, when the wavetrain amplitude approaches 0. Unstable modes in this limit should
coincide with the BTI modes that attack the homogeneous state, 
as detailed in Section 2.2. This provides a useful numerical
check on our eigensolver. In Fig.~\ref{disp1} we plot, with red dots, the growth rate
of the unstable mode that attacks 
a short-wavelength low-amplitude wavetrain with
$q=3.5724$, when $\tau_0=0.175$ and $\mu=0.025$. This wave
possesses an amplitude of $8.7\times10^{-3}$. The
solid line is the growth rate computed from the linear theory of the
homogeneous state. The agreement is excellent and thus verifies our
mathematical and numerical apparatus. 

\begin{figure}
\begin{center}
\scalebox{0.4}{\includegraphics{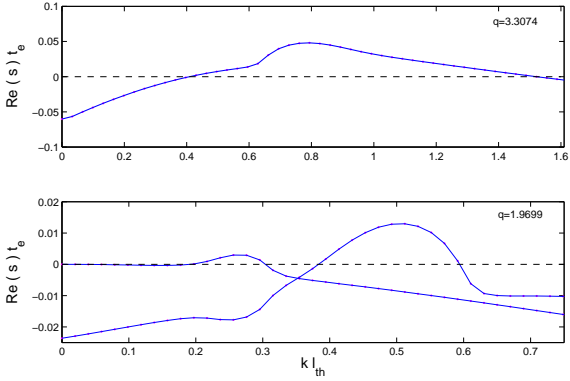}}
 \caption{The dependence on wavenumber $k$ 
  of the growth rates $s$ of secondary modes attacking two
   wavetrains. In both cases $\tau_0=0.2$ and
   $\mu=0.025$. The top panel is associated with a wavetrain with
   $q=3.3074$, the bottom panel with a wavetrain with $q=1.9699$. Note
 that the longer wave supports two growing modes.}\label{2ndstab}
\end{center}
\end{figure}

When the wavetrain amplitude becomes larger the dispersion relation deviates 
from that of the homogeneous case in Fig.~\ref{disp1}. Ultimately,
more than one mode can possess a positive growth rate as $k$ varies. 
In Fig.~\ref{2ndstab}, examples of growing modes for two different
wavetrains are shown for a low $\tau_0$ ring. The top panel describes
the sole growing mode that attacks a shorter wavelength wave. The
bottom panel presents two potentially growing modes that destabilise a
longer wave. Other wavetrains support similarly complicated dispersion
relations, which we need not go into the details of. It is important
to note that there is always a neutral mode $s=0$ when $k=0$. It is
linked to the translational invariance of our model.

Of more interest is a stability criterion in terms of $q$, for given 
$\tau_0$ and $\mu$. Essentially, which wavetrains are stable and which
are not? In general, in a given family of wavetrains we find that both
its shortest and longest members are
unstable. At low $\tau_0$ this is expected: wavetrains near the upper
and lower critical $q$'s have low amplitudes and thus will have similar
stability properties to the (unstable) homogeneous state --- cf.\
Fig.~\ref{Cringvsq}a. It is interesting that this is also the case for
larger $\tau_0$ solutions, which can have large amplitudes at lower $q$.

\begin{figure}
\begin{center}
\scalebox{0.4}{\includegraphics{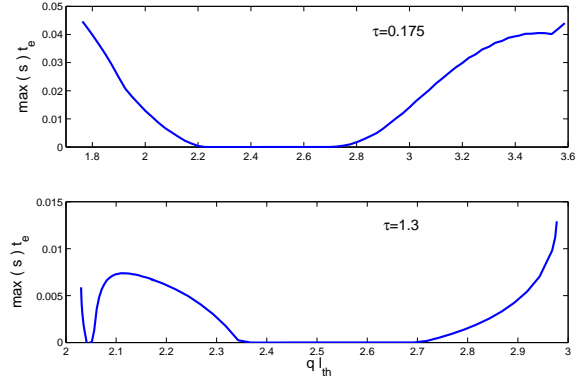}}
 \caption{The maximum growth rate $s$ of the secondary instability as
   a function of the wavenumber $q$ of the underlying wavetrain. 
  The first panel deals with wavetrains possessing $\tau_0=0.175$, and
   the second panel deals with upper-branch wavetrains possessing
   $\tau_0=1.3$. 
   The ballistic
   Prandtl number is held fixed at $\mu=0.025$. When max($s$)$ \le 0$ the
   wavetrain is stable.}\label{maxs}
\end{center}
\end{figure}

For our fiducial C-ring parameters of Section 3.2,
only wavetrains with $2.225<q<2.705$ are linearly stable. This result
is illustrated in the first panel of 
Fig.~\ref{maxs}, which shows the growth rate $s$,
maximised over $k$, as a function of wavetrain wavenumber $q$. We have
stability when
the curve takes values equal or less than 0. The stable band of
wavenumbers encompasses waves with the largest amplitude and
slowest wavespeed (see Fig.~\ref{Cringvsq}), a result that may have
been anticipated. 
The most nonlinear
solution will most effectively distort the background equilibrium
state and hence mitigate the conditions favourable for BTI.
 
Note that the fastest growing linear BTI mode
possesses a wavenumber ($q=2.84$) outside this narrow range. This
mode will dominate all others in the initial phase of a ring's
evolution, and probably saturates by forming a nonlinear wavetrain of the
same wavenumber. Being an unstable solution, however,
the
system eventually migrates away and probably undergoes a
wavelength selection process as it seeks the stable set of solutions.   
Similar behaviour is witnessed during the saturation of the viscous
overstability, with the ring hopping from one unstable solution to
another until it finds a stable wavetrain (Latter \& Ogilvie 2009,
2010). 

For B-ring parameters with $\tau_0=1.3$ we find that the lower branch
of solutions in Section 3.3 is unstable for all $q$. These waves are
destabilised by a fast growing $k=0$ mode that seems to bear little
resemblance to the classical BTI of the homogeneous state. The upper
branch, however, possesses a band of stable solutions for
$2.356<q<2.698$, which includes the waves with larger amplitudes (see Fig.~\ref{Bringvsq}). 
But there is also an unusual much narrower band of stable states around
$q=2.05$. These results are summarised in the second panel of Fig.~\ref{maxs}.

Overall this stability behaviour is reflected at other
$\tau_0$. Almost always, stable
solutions exist in a narrow band of $q$ bracketing the largest
amplitude waves.

\section{Dynamics of long and slow modulations}

So far we have uncovered the invariant fixed points of the BTI dynamical
system, i.e.\ steadily travelling wavetrains. 
These should control its nonlinear evolution, 
and we check exactly how in Paper 3 with
 numerical simulations. 
But it is possible to obtain an analytic handle on the full
time-dependent
 dynamics in certain relevant limits, especially
near marginal stability. Here we derive reduced
equations for the long and slow dynamics of the waves' modulations by
exploiting the separation of scales between the modulations and their
carrier waves. We find that the modulations of low-amplitude wavetrains
are governed by the complex Ginzburg-Landau equation (CGLE), which is a
partial differential equation that describes generic nonlinear wave
phenomena in diverse settings (Aranson and Kramer 2002). The
mathematical derivation is located in the Appendix, in this section we
briefly summarise its main points and implications for the
C-ring.

We select a point on
the curve of marginal linear stability described in Fig.~\ref{stab1}
in the parameter space of $(\tau_0, \mu)$. Next we 
move off the stability curve, either by slightly perturbing $\mu$ or $\tau_0$. 
In the Appendix,
we choose $\mu$ as it simplifies the mathematics somewhat. The perturbation's
proximity to marginality is quantified by the small dimensionless
parameter $\epsilon$. It also serves to separate the scales of the 
(fast) underlying waves and
the (slow) modulations. The former depend only on the short space and
time variables of $x$ and $t$, while the latter depend only on the long
space and time variables $X\sim \epsilon x$ and $T\sim \epsilon^2 t$. 
In this limit, to leading order, the solution behaves as
\begin{align} \label{ansatz}
\tau = \tau_0 + A(X,\,T)\text{e}^{\ii k_c x - \ii\omega_c t} + \text{c.c.},
\end{align}
where $k_c$ and $\omega_c$ are the wavenumber and frequency of the
marginal linear mode, $A$ is the complex-valued wavetrain amplitude, and `c.c.' indicates the complex conjugate. 
The amplitude $A$ describes the
modulations and obeys a version of the CGLE,
\begin{align} \label{cglet}
\d_T A = \lambda A + \beta\, |A|^2 A + \kappa\, \d_X^2 A,
\end{align}
where $\beta$ and $\kappa$ are (complex) constants, and $\lambda$ is a
control parameter. 
In the Appendix we
give expressions for $\beta$ and $\kappa$ in terms of the various
parameters of the ring. 

The set of solutions to Eq.~\eqref{cglet} includes steady homogeneous
solutions in which $A$ is a constant; these connect to the solutions
computed in Section 3. But the CGLE also admits plane waves with $A
\propto \text{exp}(\ii k_m X - \ii \omega_m T)$, where
$k_m$ and $\omega_m$ are the wavenumber and
frequency of long modulations. Some or all of
these solutions may be unstable, in which case various time-dependent
behaviours can emerge, including the aperiodic emergence and
destruction of strong inhomogeneities in the carrier
wavetrain (wherein its phase jumps abruptly) as well as low-level
chaotic variations in the waves' amplitude (Aranson and Kramer 2002). 
It is likely that low amplitude undulations in the C-ring could undergo some subset of
these dynamics.

 For the moment, we use \eqref{ansatz} and
\eqref{cglet} to compute steady wavetrains with no modulations in order
to compare with some of the results in Section 3. In Figure
\ref{Hyst-mu} the solid curve represents the real wave amplitude from
the nonlinear analysis in the Appendix, cf. Eqs \eqref{amp}-\eqref{comparison}. As expected, the
agreement is good at low amplitudes, but the two solutions deviate as
the lower branch curves upwards towards the saddle-node.

\section{Discussion}

In this final section we summarise our
results and apply them to the observational problems of
Saturn's B- and C-rings. We also point towards future work.

First, we have shown that the BTI can saturate
via the formation of steadily travelling nonlinear wavetrains.
 Near marginal
stability at low $\tau$, these solutions inhabit an interval of intermediate wavenumber $q$. For
example, when the mean optical depth $\tau_0=0.175$, wavetrains exist
with a $q$ between 1.75 and 3.59 (in units of $1/l_\text{th}$). The
amplitudes of these waves are relatively small, with $\tau$ varying
by $\sim 0.1$ between peaks and troughs (cf.\ Fig.~\ref{Cringprofs}). 
For the most part, these
solutions are close to sinusoidal in appearance and possess phase
speeds approximately equal to the linear BTI modes (Fig.~\ref{Cringvsq}).  

On the other hand, at large $\tau$ the
system permits hysteresis: even if a homogeneous ring is linearly stable
it can still support large-amplitude wavetrain solutions
via the ballistic transport mechanism.
 Consequently, the ring
will want to evolve to either the flat homogeneous state, or a stable
wave state. The wavetrains do not resemble sinusoids, and the peak to
trough variation is large, varying between 1 and 1.6 in $\tau$
(Fig.~\ref{BringProf}). Wavecrests in this marginal high-$\tau$
parameter regime propagate extremely slowly, at most with a phase speed
$\sim 0.01 \,l_\text{th}/t_e$ or 1-100 mm yr$^{-1}$ (see Fig.~\ref{Bringvsq}). 

We tested the linear stability of these structures and found that for
given parameters only a subset of the wavetrain solutions are stable. 
Generally, stable solutions possess the greatest amplitude
and propagate the slowest. It is likely that the system will
select one of these
solutions if left to freely evolve.
We also demonstrate that low-amplitude wavetrains undergo large-scale 
modulations which are governed by
the complex Ginzburg-Landau equation. The amplitudes of 
our C-ring waves may then
share in its interesting, sometimes disordered, dynamics.

These results are compatible with observations of B- and C-ring
structure (Porco et al.~2005, Colwell et al.~2009), 
as well as previous simulations of the inner B-ring (D92). 
Turning to the B-ring first, it is likely that the observed
adjoining flat and wave zones between 93,000 and 98,000 km (Fig.~13.13
in Colwell et al.~2009) 
are products of the hysteresis exhibited
by our model.
The flat
regions correspond to where the ring has fallen into the stable
homogeneous state, and the wave regions to where it has jumped into
the stable wave state. These zones are connected by fronts, which
should exhibit additional dynamics that time-dependent simulations may
probe. Perturbations that may have thrust
B-ring regions out of the
homogeneous state's basin of attraction might include the
inner B-ring edge, the transition to extremely large $\tau$ at
$r=99,000$ km, or the Janus/Epimetheus 2:1 inner Lindblad
resonance. 

There are two problems that this scenario faces. 
First is the deepness of the troughs
in the theoretical wave profiles. Typically 
the theoretical troughs possess a dynamical
optical depth of $\sim 0.4$.
Meanwhile in the B-ring the troughs yield a photometric optical depth
of $\sim 0.8$. 
It is true that self-gravity wakes complicate the relationship between
dynamical and photometric optical depth, yet the discrepancy is
concerning. Second, is the observed mean optical depth differs in the
wave and in the flat zones: in the former it is approximately 0.8; in
the latter is is closer to 1.3. This further complicates the mapping of our results
to the observations, and indicates our theory requires additional refinement. 

Comparison of our solutions to C-ring observations must first resolve
one key question: does the free evolution of the BTI generate the 
low-amplitude 1000-km undulations, found between 77,000 km and 86,000 km, 
or the larger-amplitude 100-km plateaus, between 84,000 and 91,000 km
(Fig.~13.17 in Colwell et al.~2009)? 
On account of the small amplitudes and morphology of our wavetrain
solutions, we conclude that the 1000-km undulations are the
result of the BTI working alone. The plateaus
are probably caused by something else, though
the ballistic transport process may influence their
general shape (Estrada \& Durisen 2010).

If both the 1000-km undulations in the C-ring
and the 100-km waves in the B-ring are
BTI wavetrains then it follows
that $l_\text{th}$ could vary significantly between the two
radial locations. This variation may arise from differences in the
sizes, composition, or regolith properties of the ring particles, on
the one hand, or the trajectories
and speeds of the incoming meteoroids, on the other. For example,
recent spectroscopic studies indicate that the sizes of regolith
grains vary with radius (Morishima et al.~2012, Hedman et al.~2013).
But it is unclear whether this means particles are more or
less `fluffy' (and hence $l_\text{th}$ smaller or greater) in
different ring regions.
We view this as a key question in the
study of ballistic transport, deserving of further study\footnote{Note that the recent impacts
  observed by Tiscareno et al.~(2013) involved cm to m sized
  meteoroids and, being in a different collisional regime, cannot help
  constrain $l_\text{th}$.}.

In our following paper, the role of these invariant solutions is made
clear through full time-dependent simulations. There we also run a
suite of simulations of the inner B-ring edge, which itself could be
unstable to the BTI. Further work will improve our basic model,
through the addition of more physical processes. 
For instance, the ring's viscosity should be an increasing
function of $\tau$, not a constant as assumed here. Preliminary
results, however, show no qualitative changes arises from this
effect. Of greater importance may be the form of the absorption
probability, $P$. Throughout this paper, we assume it only depends on
the absorbing radius. But in lower optical depth regions it will also
depend on the ejecta emitting radius. This effect will influence both
the B and C-rings, the former on account of the low $\tau$ achieved in
wavetrain troughs.

\section*{Acknowledgments}
The authors would like to thank Dick Durisen and the reviewer, Sebastien Charnoz, for
helpful comments. This research was supported by STFC grants ST/G002584/1 and ST/J001570/1.

\appendix

\section{Weakly nonlinear analysis}

\begin{figure}
\begin{center}
\scalebox{0.5}{\includegraphics{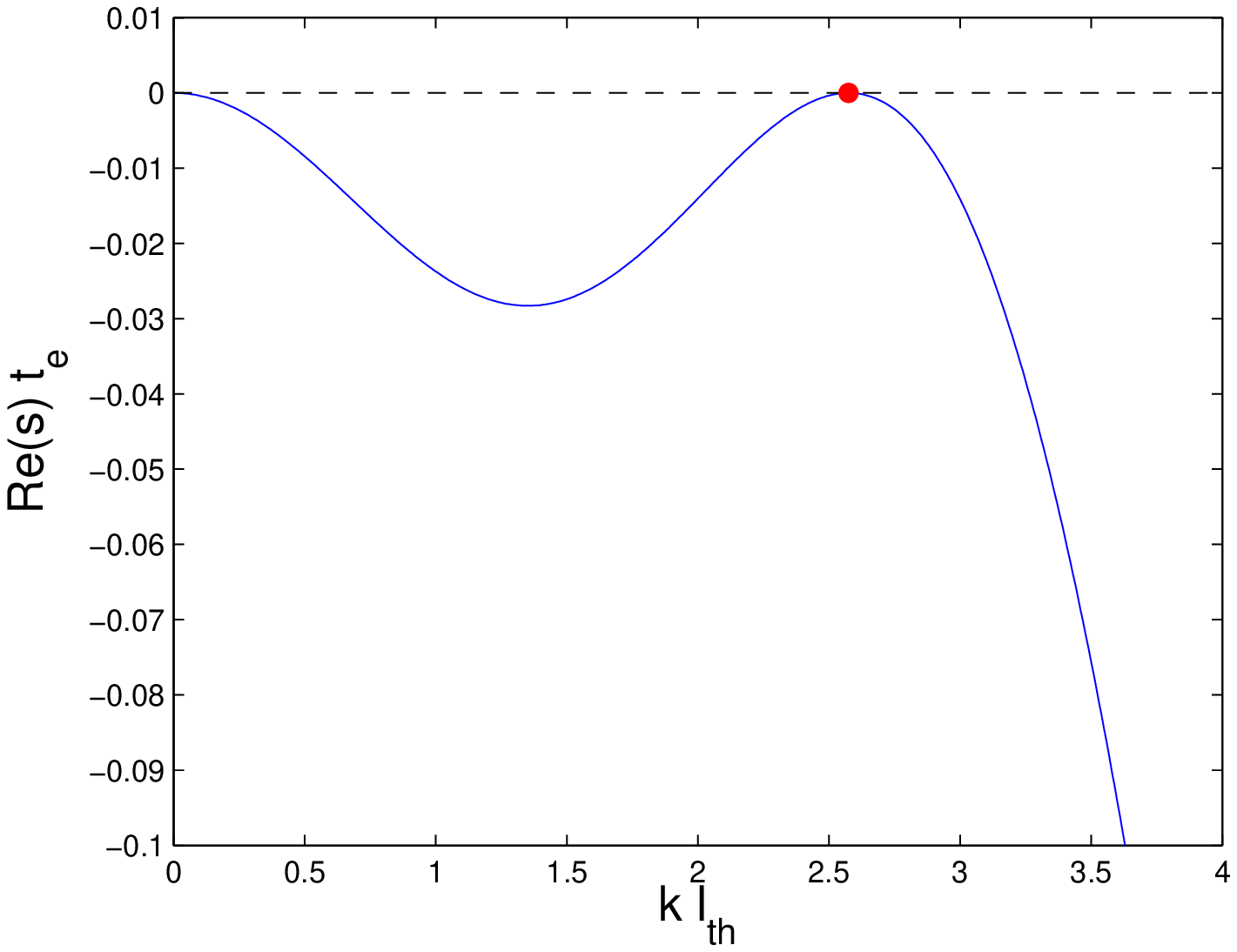}}
 \caption{The linear dispersion relation of the BTI for a marginal
   case. Here the example is taken of $\tau=1$ and
   $\mu=0.0376287$. The critical $k$ in this situation is
   $k=2.57$, which is illustrated with a red dot.}\label{crit}
\end{center}
\end{figure}

\subsection{Critical state}

We first define the critical state for which the BT instability has
zero growth rate for non-zero $k$. In Fig.~\ref{crit} we plot the dispersion
relation of the BTI at marginality when $\tau_0=1$ and $\mu\approx 0.0376$. 
Formally, a general marginal state is defined via
\begin{align}
\text{Re}(s)=0, \qquad \text{Re}\left(\frac{ds}{dk}\right)=0.
\end{align}
We set $\tau_0$ as a free parameter,
 and solve these two equations for the critical $\mu$ and $k$,
hence denoted by $\mu_c$ and $k_c$. A linear mode with the latter wavenumber
has zero growth rate but non-zero wave frequency
$\omega_c=\omega(k_c)$. We define the linear wave frequency to be
$\omega= -\text{Im}(s)$. The phase
speed is hence $c_p= \omega_c/k_c$ and its group velocity is $c_g =
(d\omega/dk)_c$.

\subsection{Slow variables and expansions}

We introduce the small dimensionless parameter $\epsilon$, so
that $0<\epsilon\ll 1$. The critical $\mu$ is then perturbed very
slightly so that
\begin{align}
\mu = \mu_c - \epsilon^2\lambda/k_c^2,
\end{align}
where $\lambda$ is a control parameter describing the proximity of the
ring to marginality. When $\lambda<0$ the system is subcritical, and
when it is $>0$ the system is supercritical.
Next we consider the long space and slow time
variables 
\begin{align}
X = \epsilon (x- c_g t), \qquad T = \epsilon^2 t.
\end{align}
If $\tau$ depends independently on $x$, $t$, $X$,
and $T$, we may replace the partial derivatives in our governing
equation as follows:
\begin{align}
\d_x \to \d_x + \epsilon\, \d_X, \qquad \d_t \to \d_t -\epsilon\, c_g\, \d_X
+ \epsilon^2 \d_T.
\end{align}

We expand $\tau$ in small $\epsilon$ around the reference
optical depth $\tau_0$:
\begin{align}
\tau = \tau_0 + \epsilon\,\tau_1 + \epsilon^2\,\tau_2 + \epsilon^3\,\tau_3+\dots.
\end{align}
Correspondingly we expand
both $R$ and $P$, obtaining
\begin{align}
P &= P_0 + \epsilon\,P_0'\,\tau_1 + \epsilon^2\left(P_0'\tau_2+
  \tfrac{1}{2}P_0''\tau_1^2\right) \notag \\
  & \hskip1cm \epsilon^3\left(P_0'\tau_3 + P_0''\tau_1\tau_2 + \tfrac{1}{6}P_0'''\tau_1^3\right)+\dots,
\end{align}
where the subscript $0$ indicates evaluation at $\tau=\tau_0$ and a
prime indicates differentiation with respect to $\tau$. An analogous
expression exists for $R$. 

\subsubsection{Direct mass transfer integrals}

Consider first the integral operator 
\begin{align}
\mathcal{I}= P[\tau(x)]\,\int R[\tau(x-\xi)]\, f(\xi)\, d\xi.
\end{align}
On placing the above expansion into the integral we are faced with
integrals of the form
\begin{align}\label{I1}
 \int h(x-\xi) f(\xi) d\xi,
\end{align}
where $h$ is a nonlinear combination of the $\tau_i$. 

According to the scale separation we treat $h$ as a function of both
$x$ and $X$. At any given instant $t$ we 
replace $h(x-\xi)$ by $h(x-\xi,\,X-\epsilon
\xi)$ in \eqref{I1}. Next $h$ is expanded as a Taylor series in its
second argument,
\begin{align}
h(x-\xi,\,X-\epsilon\xi)= 
\sum_{n=0}^{\infty} \epsilon^n (1/n!) (-\xi)^n (\d_X)^n h(x-\xi,\,X).
\end{align}
Expression \eqref{I1} then becomes
\begin{align}
 \sum_{n=0}^{\infty}
\epsilon^n I_n\left(\d_X^n h\right),
\end{align}
where we have introduced the following family of integral operators
\begin{align}
I_n(h) = \frac{1}{n!}\int (-\xi)^n\, h(x-\xi,X)\, f(\xi)\, d\xi.
\end{align}

We can do the same with the $\mathcal{J}$ operator, which throws up terms
such as 
\begin{align}
\int h(x+\xi)\,f(\xi)\,d\xi = \sum_{n=0}^\infty \epsilon^n
J_n\left(\d_X^n h\right),
\end{align}
where
\begin{align}
J_n(h) = \frac{1}{n!}\int \xi^n\, h(x+\xi,X)\, f(\xi)\, d\xi.
\end{align}

Putting all this together and collecting orders of $\epsilon$ 
we have the following expansions:
\begin{align}
\int R[\tau(x-\xi)] f(\xi) d\xi & = M_0 + \epsilon M_1 + \epsilon^2 M_2
+ \epsilon^3 M_3 + \dots \\
\int P[\tau(x+\xi)] f(\xi) d\xi & = N_0 + \epsilon N_1 + \epsilon^2 N_2
+ \epsilon^3 N_3 + \dots,
\end{align}
where
\begin{align}
M_0 &= R_0, \\
M_1 &= R_0'\,I_0(\tau_1), \\
M_2 &= R_0'[I_1(\d_X\tau_1) +I_0(\tau_2)] + \tfrac{1}{2}R_0''
I_0(\tau_1^2), \\
M_3 &= R_0'[I_2(\d_X^2\tau_1)+ I_1(\d_X\tau_2)+I_0(\tau_3)] \notag\\ 
 & +R_0''[\tfrac{1}{2}I_1(\d_X\tau_1^2)+I_0(\tau_1\tau_2)] +
\tfrac{1}{6}R_0''' I_0(\tau_1^3),
\end{align}
and
\begin{align}
N_0 &= P_0, \\
N_1 &= P_0'\,J_0(\tau_1), \\
N_2 &= P_0'[J_1(\d_X\tau_1) +J_0(\tau_2)] + \tfrac{1}{2}P_0''
J_0(\tau_1^2), \\
N_3 &= P_0'[J_2(\d_X^2\tau_1)+ J_1(\d_X\tau_2)+J_0(\tau_3)]\notag \\ 
 & +P_0''[\tfrac{1}{2}J_1(\d_X\tau_1^2)+J_0(\tau_1\tau_2)] +
\tfrac{1}{6}P_0''' J_0(\tau_1^3).
\end{align}

For reference, the operation of the $I_n$ and $J_n$ on plane waves
gives:
\begin{align}
I_n(\text{e}^{\ii kx}) = \frac{1}{n!}\,(-\ii)^n\,F^{(n)}(k)\text{e}^{\ii kx}, \\
J_n(\text{e}^{\ii kx}) = \frac{1}{n!}\,(\ii)^n\,F^{(n)}(-k)\text{e}^{\ii kx}.
\end{align}

\subsubsection{Angular momentum transfer integrals}

The integral operators $\mathcal{K}$ and $\mathcal{L}$ 
associated with the angular momentum terms are
treated similarly. We derive the following expansions:
\begin{align}
\int R[\tau(x-\xi)]\xi\,f(\xi)\,d\xi &= U_0 + \epsilon U_1 + \epsilon^2
U_2 + \epsilon^3 U_3 + \dots \\
\int P[\tau(x+\xi)]\xi\,f(\xi)\,d\xi &= V_0 + \epsilon V_1 + \epsilon^2
V_2 + \epsilon^3 V_3 + \dots.
\end{align}
Here 
\begin{align}
U_0 &= i\,R_0\,F'(0), \\
U_1 &= R_0'\,K_0(\tau_1), \\
U_2 &= R_0'[K_1(\d_X\tau_1)+ K_0(\tau_2)]+\tfrac{1}{2}R_0''
K_0(\tau_1^2), \\
U_3 &= R_0'[K_2(\d_X^2\tau_1) + K_1(\d_X\tau_2)+ K_0(\tau_3)] \notag \\
    & + R_0''[\tfrac{1}{2}K_1(\d_X\tau_1^2)+K_0(\tau_1\tau_2)] + \tfrac{1}{6}R_0'''K_0(\tau_1^3),
\end{align}
where the $K_i$ are defined through
\begin{align}
K_n(h) = \frac{1}{n!}\int (-1)^n \xi^{n+1}\,h(x-\xi,X)\,f(\xi)\,d\xi,
\end{align}
and
\begin{align}
V_0 &= i\,P_0\,F'(0), \\
V_1 &= P_0'\,L_0(\tau_1), \\
V_2 &= P_0'[L_1(\d_X\tau_1)+ L_0(\tau_2)]+\tfrac{1}{2}P_0''
L_0(\tau_1^2), \\
V_3 &= P_0'[L_2(\d_X^2\tau_1) + L_1(\d_X\tau_2)+ L_0(\tau_3)] \notag
\\
    & + P_0''[\tfrac{1}{2}L_1(\d_X\tau_1^2)+L_0(\tau_1\tau_2)] + \tfrac{1}{6}P_0'''L_0(\tau_1^3),
\end{align}
where the $L_i$ are defined through
\begin{align}
L_n(h) = \frac{1}{n!}\int \xi^{n+1}\,h(x+\xi,X)\,f(\xi)\,d\xi.
\end{align}
Note that
\begin{align}
K_n(\text{e}^{\ii kx}) &=
-\frac{1}{n!}(-\ii)^{n+1}\,F^{(n+1)}(k)\,\text{e}^{\ii kx}, \\
L_n(\text{e}^{\ii kx}) &=
\frac{1}{n!}\ii^{n+1}\,F^{(n+1)}(-k)\,\text{e}^{\ii kx}.
\end{align}

\subsection{Balances}

We are now in a position to establish the various orders of Eq.~(1). 

\subsubsection{Order $\epsilon$}
To leading order we obtain $\mathcal{Z}(\tau_1)=0$, where
\begin{align}
\mathcal{Z}(\tau_1)  &= \d_t \tau_1 - P_0\,M_1 - M_0 \,P_0' \tau_1 + R_0
N_1 + R_0' N_0 \tau_1 \notag \\
 & \hskip 1cm - \tfrac{1}{2}P_0' U_0\, \d_x \tau_1 -
\tfrac{1}{2}P_0\d_x U_1 - \tfrac{1}{2}R_0' V_0 \d_x \tau_1 \notag \\
& \hskip 2cm - \tfrac{1}{2}R_0 \d_x V_1 - \mu_c \d_x^2 \tau_1.
\end{align}
This is a linear equation for $\tau_1$ in the variables $t$ and
$x$. It admits (by construction) solutions of the form
\begin{align}
\tau_1 = A(X,\,T)\,\text{e}^{\ii k_c x - \ii\omega_c t}.
\end{align}
So at this order the solution is the critical linear BTI mode with a complex amplitude $A$
that depends on the slow variables. We now define
$$ a(x,t) = \text{e}^{\ii k_c x - \ii\omega_c t},$$
and take the general solution at this order to be
\begin{align}
\tau_1 = A(X,T)\,a(x,t) + \overline{A(X,T)}\,\overline{a(x,t)}.
\end{align}

\subsubsection{Order $\epsilon^2$}

At next order, after considerable algebra, we obtain the
following for $\tau_2$:
\begin{align}\label{order2}
\mathcal{Z}(\tau_2)= G(k)\,A^2a^2+ \overline{G(k)}\,\overline{A^2a^2},
\end{align}
where
\begin{align}
G(k)=& \tfrac{1}{2}P_0\,R_0'' H(2k)
 -\tfrac{1}{2}R_0P_0''\overline{H(2k)}
\notag\\
&+P_0'R_0'\left\{F(k)-\overline{F(k)}-k\left[F'(k)+\overline{F'(k)}\right] \right\}.
\end{align}
In the above we have dropped the subscript $c$ on $k_c$.
Note that there are no terms on the right side of \eqref{order2} that
are linear in $a$; we are then assured that the equation is solvable
for $\tau_2$. 

We assume a solution of the form
\begin{align}
\tau_2 = B(X,T)a^2 + \overline{B(X,T)}\,\overline{a^2},
\end{align}
where $B$ is a complex amplitude to be determined.
Using the fact that
\begin{align}
\mathcal{Z}(a^2)=-[2\ii\omega_c+s(2k)]a^2,
\end{align}
we obtain
\begin{align}
B = -\frac{G(k)}{2\ii\omega_c+ s(2k)}\,A^2.
\end{align}

\subsubsection{Order $\epsilon^3$}

The equation at next order can be put in the following form
\begin{align}
\mathcal{Z}(\tau_3)= Z_1\,a +Z_2\, a^2 + Z_3\,a^3 + \text{c. c.},
\end{align}
in which the $Z_i$ are complicated expressions. In order to solve this
equation we require $Z_1=0$, because $a^2$, $a^3$, etc are
orthogonal to $a$. 
This equation is a version of the CGLE
 for the
complex amplitude
$A$:
\begin{align} \label{cgle}
\d_T A = \lambda\,A + \beta\,A|A|^2 + \kappa\,\d_X^2 A.
\end{align}
Here the diffusion coefficient is
\begin{align}
\kappa &= \tfrac{1}{4}P_0 R_0'\,k\,F'''(k) + \tfrac{1}{4}R_0 P_0'\,k
F'''(-k)+\mu_c, \\
       &= -\frac{1}{2}\frac{\d^2 s}{\d k^2},
\end{align}
in line with expectations from the linear dispersion relation.
The coefficient of the nonlinear term is much more involved,
\begin{align}
\beta &= \tfrac{1}{2}\left[P_0 R_0''' H(k)- R_0 P_0''' \overline{H(k)}\right]
\notag \\
      & \hskip1cm +\tfrac{1}{2}\left[P_0' R_0'' C(k) - R_0'P_0'' \overline{C(k)} \right]
      \notag \\
      & \hskip1cm -\frac{G(k)\left[P_0 R_0'' H(k)-R_0 P_0'' \overline{H(k)} + P_0' R_0' D(k)\right]}{2\ii\omega_c+s(2k)}.
\end{align}
Here we have introduced the following functions of $k$:
\begin{align}
C(k) &= F(2k)-F(k) + 2 F(0)-2F(-k) \notag \\ 
     & \hskip0.5cm -
     \tfrac{1}{2}k\left[2F'(0)+2F'(-k)+F'(2k)+F'(k)\right], \\
D(k) &= F(2k)-F(-2k) - \tfrac{1}{2}k\left[F'(2k)+F'(-2k)\right] \notag
\\
     & \hskip0.5cm +F(-k)-F(k) - \tfrac{1}{2}k\left[F'(-k)+F'(k)\right].
\end{align}

\subsection{Plane wave modulations}
Equation \eqref{cgle} admits plane wave solutions:
\begin{align}
A = |A| \text{exp}\left(\ii k_m X - \ii \omega_m T\right),
\end{align}
where $k_m$ and $\omega_m$ are the wavenumber and (real) frequency of the
amplitude modulation. The wavenumber $k_m$ is a free parameter, and
the frequency can be determined from the `nonlinear dispersion
relation'
\begin{align}
\omega_m = -\beta_i |A|^2 + k_m^2\,\kappa_i,
\end{align}
where the subscript $i$ indicates imaginary part. The amplitude of the
wave is set by the real part of \eqref{cgle} divided by $A$,
\begin{align} \label{amp}
|A|^2 = \frac{k_m^2\kappa_r-\lambda}{\beta_r},
\end{align}
where the subscript $r$ indicates real part. The unmodulated
wavetrains computed in Section 3 have $k_m=0$. These solutions can
then be written in a form more convenient for the comparison in
Fig.~9, 
\begin{align} \label{comparison}
\tau \approx \tau_0 + 2\sqrt{\frac{\mu-\mu_c}{\beta_r}}\,\cos\left(k_c x -\omega_c t\right).
\end{align}


\begin{thebibliography}{40}

\bibitem{ArKr}
Aranson, I.O., Kramer, L., 2002. RvMP, 74, 99.

\bibitem{ArT}
Araki, S., Tremaine, S., 1986.
Icarus, 65, 83.

\bibitem{Boyd}
Boyd, J.P., 2002. Chebyshev and Fourier Spectral Methods, 2nd ed,
Dover Press, New York.

\bibitem{BurkeKnob}
Burke, J., Knobloch, E., 2007. Chaos, 17, 037102.

\bibitem{Ch}
Charnoz, S., Dones, L., Esposito, L.W., Estrada, P.R., Hedman, M.M., 2009.
In: Dougherty, M.~K., Esposito, L.~W., Krimigis, S.~M. (eds.),
 \emph{Saturn
  from Cassini-Huygens}, Springer, Dordrecht Netherlands, p537.

\bibitem{Col2}
Colwell, J.~E., Nicholson, P.~D., Tiscareno M.~S., Murray, C.~D., French,
R.~G., Marouf, E.~A., 2009.
In: Dougherty, M.~K., Esposito, L.~W., Krimigis, S.~M. (eds.),
 \emph{Saturn
  from Cassini-Huygens}, Springer, Dordrecht Netherlands, p375.

\bibitem{CD90}
Cuzzi, J.~N., Durisen, R.~H., 1990. Icarus, 84, 467. (CD90)

\bibitem{Dai}
Daisaka, H., Tanaka, H., Ida, S., 2001. Icarus, 154, 296.

\bibitem{D84}
Durisen, R.~H., 1984. In: Greenberg, R., Brahic, A., (Eds),
\emph{Planetary Rings}, University if Arizona Press, Tucson, p416.

\bibitem{D95}
Durisen, R.~H., 1995. Icarus, 115, 66. (D95)

\bibitem{D89}
Durisen, R.~H., Cramer, N.~L., Murphy, B.~W., Cuzzi, J.~N., Mullikin,
T.~L., Cederbloom, S.~E., 1989. Icarus, 80, 136. (D89)

\bibitem{D92}
Durisen, R.~H., Bode, P.~W., Cuzzi, J.~N., Cederbloom, S.~E., Murphy,
B.~W., 1992. Icarus, 100, 364. (D92)

\bibitem{EstDur}
Estrada, P., Durisen, R., 2010. 41st Lunar and Planetary Science
Conference Abstracts, p2686.

\bibitem{Hedman}
Hedman, M.~M., Nicholson, P.~D., Cuzzi, J.~N., 
Clark, R.~N., Filacchione, G., Capaccioni, F., Ciarniello, M., 2013.
Icarus, 223, 105. 

\bibitem{HC}
Horn, L., Cuzzi, J., 1996. Icarus, 119, 285.

\bibitem{Ip84}
Ip, W.-H, 1984. Icarus, 60, 547.

\bibitem{Lat09}
Latter, H.~N., Ogilvie, G.~I., 2009. Icarus, 202, 565.

\bibitem{Lat10}
Latter, H.~N., Ogilvie, G.~I., 2010. Icarus, 210, 318.

\bibitem{Lat12}
Latter, H.~N., Ogilvie, G.~I., Chupeau, M., 2012. MNRAS, 427,
2336. (Paper 1.)

\bibitem{Lat13}
Latter, H.~N., Ogilvie, G.~I., Chupeau, M., 2013. MNRAS,
submitted. (Paper 3.)

\bibitem{Liss84}
Lissauer, J.~J., 1984. Icarus, 57, 63.

\bibitem{Morishima}
Morishima, R., Edgington, S.~G., Spilker, L., 2012.
Icarus, 221, 888.

\bibitem{Science}
Porco, C.~C. and 34 colleagues, 2005.
Science, 307, 1226

\bibitem{Tisc}
Tiscareno, M.~S., Mitchell, C.~J., Murray, C.~D., Di Nino, D., Hedman,
M.~M., Schmidt, J., Burns, J.~A., Cuzzi, J.~N., Porco, C.~C., Beurle,
K., Evans, M.~W., 2013. Science, 340, 460.

\bibitem{WhitWat}
Whitaker, E.T., Watson, G.N., 1990. A Course in Modern Analysis, 4th
ed., Cambridge University Press, Cambridge UK.

\bibitem{WisT}
Wisdom, J., Tremaine, S., 1988.
The Astronomical Journal, 95, 925.

\end{thebibliography}
\end{document}